# SLC Final Performance and Lessons[*]

Nan Phinney[#]

SLAC, P.O. Box 4349, Stanford CA 94309  (USA)

*Abstract*

The Stanford Linear Collider (SLC) was the first prototype of a new type of accelerator, the electron-positron linear collider. Many years of dedicated effort were required to understand the physics of this new technology and to develop the techniques for maximizing performance. Key issues were emittance dilution, stability, final beam optimization and background control. Precision, non-invasive diagnostics were required to measure and monitor the beams throughout the machine. Beam-based feedback systems were needed to stabilize energy, trajectory, intensity and the final beam size at the interaction point. A variety of new tuning techniques were developed to correct for residual optical or alignment errors. The final focus system underwent a series of refinements in order to deliver sub-micron size beams. It also took many iterations to understand the sources of backgrounds and develop the methods to control them. The benefit from this accumulated experience was seen in the performance of the SLC during its final run in 1997-98. The luminosity increased by a factor of three to $3*10^{30}$ and the 350,000 Z data sample delivered was nearly double that from all previous runs combined.

## 1   INTRODUCTION

The concept of an electron-positron linear collider was proposed as a way of reaching higher energy than was feasible with conventional storage ring technology. The SLC, built upon the existing SLAC linac, was intended as an inexpensive way to explore the physics of the $Z^0$ boson while demonstrating this new technology [1]. Both goals were much more difficult to achieve than anticipated, with the SLC only approaching design luminosity after ten years of operation. As the first of an entirely new type of accelerator, the SLC required a long and continuing effort to develop the understanding and techniques required to produce a working linear collider. Precision diagnostics, feedback, automated control and improved tuning algorithms were key elements in this progress. In parallel, there was an international collaborative effort to design an $e^+e^-$ collider to reach an energy of 1 Tev or higher [2]. Both projects benefited from a close interaction. The SLC drew on the ideas and techniques developed for a future machine while the collider design has been heavily influenced by the experience gained with the SLC.

___________________________
[*]Work supported by the U.S. Dept. of Energy under contract DE-AC03-76SF00515

[#] Email: nan@slac.stanford.edu

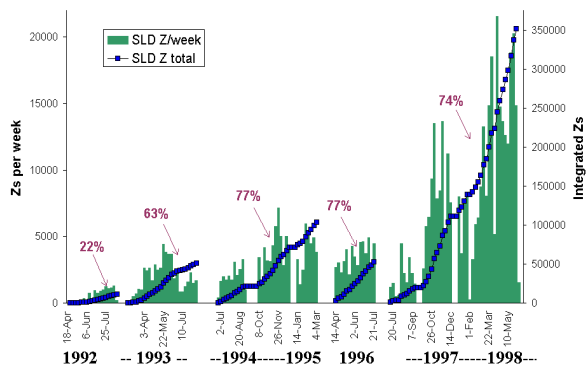

Figure 1:  SLD luminosity showing the performance improvement from 1992-1998. The bars show luminosity delivered per week and the lines show integrated luminosity for each run. The numbers give average polarization.

## 2   SLC HISTORY

The SLC was first proposed in the late 1970s, with design studies and test projects starting soon after. Construction began in October, 1983 and was completed in mid-1987, with many upgrades in succeeding years. After two difficult years of commissioning, the first $Z^0$ event was seen by the MARK II detector in April 1989. The MARK II continued to take data through 1990. In 1991 the SLD experiment was brought on line with a brief engineering run. SLD physics data taking began the next year with a polarized electron beam. More than 10,000 $Z^0$'s were recorded with an average polarization of 22%. In 1993, the SLC began to run with 'flat beam' optics with the vertical beam size much smaller than the horizontal, unlike the original design where the beam sizes were nearly equal [3]. This provided a significant increase in luminosity and SLD logged over 50,000 $Z^0$s. The polarized source had also been upgraded to use a 'strained lattice' cathode which provided polarization of about 62% [4].

  For the 1994-95 run, a new vacuum chamber was built for the damping rings to support higher beam intensity [5] and the final focus optics was modified to produce smaller beams at the Interaction Point (IP) [6]. A thinner strained lattice cathode brought the polarization up to nearly 80%. Over 100,000 $Z^0$'s were delivered in this long run. For the next runs, the SLD experiment was upgraded with an improved vertex detector with better resolution and larger acceptance. In 1996, operations were limited by scheduling constants and 50,000 $Z^0$s were delivered. The final run of the

SLC began in 1997 and continued through mid-1998. The luminosity increased by more than a factor of three and a total of 350,000 $Z^0$s were recorded, nearly double the total sample of events from all previous SLD runs [7]. Because of the high electron beam polarization, the small and stable beam size at the interaction point, and a high-precision vertex detector, the SLD was able to make the world's most precise measurements of many key electroweak parameters with this data sample. Figure 1 shows the SLC luminosity history.

## 3  1997-98 PERFORMANCE

During the 1997-98 run, the SLC reached a peak luminosity of 300 $Z^0$s per hour or $3*10^{30}$ /cm$^2$/sec. The luminosity steadily increased throughout the run, demonstrating that the SLC remained on a steep learning curve. A major contribution to this performance came from a significant disruption enhancement, typically 50-100%. The improvement was due to changes in tuning procedures and reconfiguration of existing hardware with no major upgrade projects. Improved alignment and emittance tuning procedures throughout the accelerator resulted in minimal emittance growth from the damping rings to the final focus. In particular, a revised strategy for wakefield cancellation using precision beam size measurements at the entrance to the final focus proved effective for optimizing emittance. The final focus lattice was modified to provide stronger demagnification near the interaction point and to remove residual higher-order aberrations.

The luminosity of a linear collider $L$ is given by

$$L = \frac{N^+ N^- f}{4\pi \sigma_x \sigma_y} H_d \quad (1)$$

where $N^{\pm}$ are the number of electrons and positrons at the interaction point (IP), $f$ is the repetition frequency, $\sigma_{x,y}$ are the average horizontal ($x$) and vertical ($y$) beam sizes, and $H_d$ is a disruption enhancement factor which depends on the beam intensities and the transverse and longitudinal beam sizes. At the SLC, the repetition frequency was 120 Hz and the beam intensity was limited by wakefield effects and instabilities to about $4*10^{10}$ particles per bunch. The only route to higher luminosity was by reducing the effective beam size. Taking emittance as the product of the beam size and angular divergence ($\theta_{x,y}$), $\varepsilon_{x,y} = \sigma_{x,y} \theta_{x,y}$, the basic strategy was to decrease the emittance and increase the angular divergence. A key breakthrough was the understanding that the effective beam size, $\sigma_{x,y}$, must be evaluated from the integral over the beam overlap distribution and not the RMS. Properly calculated, $\sigma_{x,y}$ decreases with larger $\theta_{x,y}$ as shown in Figure 2. Further reduction of the vertical size was possible by the addition of a permanent magnet octupole on each side of the final focus as shown in Figure 3.

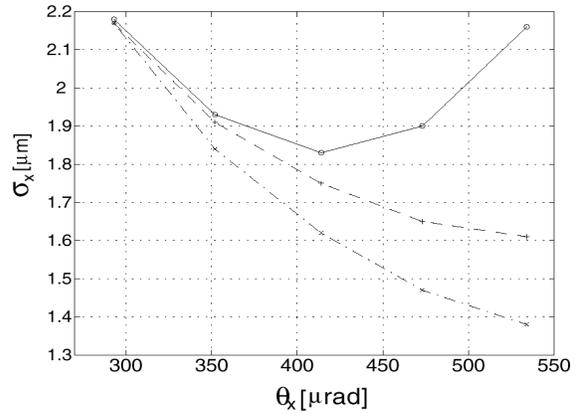

Figure 2: Horizontal beam size vs angular divergence at the SLC IP showing the reduction in beam size for larger $\theta^*$. The upper curves are for the 1996 optics, calculated using the RMS beam size (solid) and correct luminosity-weighted effective beam size (dashed). The lower curve (dot-dashed) is for the 1998 optics.

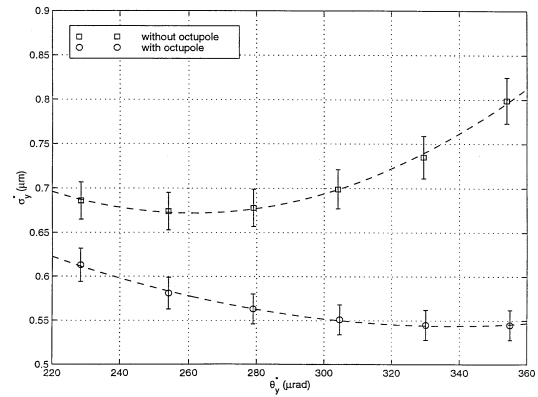

Figure 3: Vertical beam size vs angular divergence at the SLC IP showing the dependence of beam size $\sigma_y$ on divergence $\theta_y$, without (upper) and with octupoles (lower) to cancel higher order aberrations. Both curves are the luminosity-weighted effective beam size.

Beam sizes as small as 1.5 by 0.65 microns were achieved at full beam intensity of $4*10^{10}$ particles per pulse. With these parameters, the mutual focussing of the beams in collision becomes significant, resulting in a further increase in luminosity. The strength of the effect is characterized by the disruption parameter, $D_{x,y}$, for each plane which is the inverse focal length in units of the bunch length, $\sigma_z$.

$$D_{x,y} = \frac{2 N r_e \sigma_z}{\gamma \sigma_{x,y}(\sigma_x + \sigma_y)} \quad (2)$$

Recorded SLD event rates confirmed the theoretical calculations of the disruption enhancement which was typically 50-100% [8]. Figure 4 shows the measured disruption enhancement. The enhancement is calculated as the ratio of the measured SLD event rate to that predicted for rigid beams without disruption.

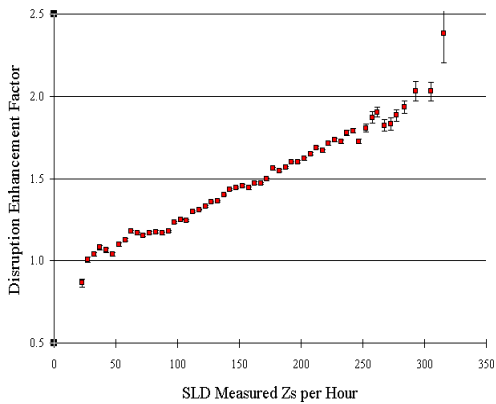

Figure 4: Measured disruption enhancement factor as a function of luminosity. At the highest luminosity, the enhancement exceeded 100%.

## 4 DIAGNOSTICS

A key element in improving the performance of the SLC was the development of precision, non-invasive diagnostics to characterize and monitor the beams. Breakthroughs in understanding often followed quickly on the heels of a new diagnostic tool which allowed insight into the beam quality and correlations. One example was the characterization of a microwave instability in the damping rings. Beginning in 1990, the maximum beam intensity was limited by errant pulses which created high backgrounds in the detector. By correlating the energy and trajectory on a pulse-to-pulse basis, the problem was traced to the damping rings. Only when a diagnostic was developed to monitor the bunch length continuously while the beam was in the rings was it possible to identify the cause, a longitudinal instability due to the interaction of the intense bunches with the impedance of the vacuum chambers. The beam intensity could be increased only after these chambers were rebuilt in 1994.

Emittance preservation in a linear collider requires tight control of the trajectories and optical matching in the linacs and transport lines. If the beam is not matched to the lattice at the entrance of the linac, the inherent energy spread of the beam will cause slices of different energy to filament. Dispersion from the beam passing off-axis through the quadrupoles interacts with the correlated energy spread along the beam to create an $x$–$z$ (or $y$–$z$) correlation or tilt. If the beam passes off-axis through the structures, wakefields from the head of the bunch act on the tail to cause another $x$–$z$ correlation. To avoid these effects, one must be able to accurately characterize the beam profile and optimize it as a function of lattice and trajectory changes.

The key to emittance control at the SLC was the development of wire scanners which allowed a precise, rapid, non-invasive measurement of the beam profile. The first scanners were installed at the beginning and end of the linac in 1990 [9]. Four scanners separated in betatron phase provide a measurement of the beam emittance in a few seconds. The wires scan across the beam during a sequence of pulses scattering a small fraction of the particles on each pulse. Downstream detectors measure the number of scattered particles at each step to map out the beam profile. Wire scanners were absolutely essential for matching the positron beam into the SLC linac since invasive monitors like fluorescent screens would interrupt the electrons needed to produce more positrons. Over several years, more than 60 wire scanners were installed throughout the SLC from the injector to the final focus to characterize the beam transverse size and energy distribution. Many of these were scanned routinely by completely automated procedures to provide real-time monitoring and long term histories of the beam properties. In 1996, a novel 'laser wire' beam size monitor was developed and installed near the SLC IP [10] to measure the individual micron-scale beams which would destroy any conventional wire. This device placed an optical scattering center inside the beam pipe with light from a high power pulsed laser brought to a focus of 400-500 nm. The $e^+$ or $e^-$ beam was scanned across the laser spot and its shape reconstructed from the number of scattered particles at each step. This device was a prototype for the beam size monitors which will be needed for the micron-size beams of a future linear collider.

To maintain the optical matching to high precision required not only the development of the measurement devices themselves but many iterations of refinements in the data processing algorithms. Typically four wires were used to provide a redundant measurement of the phase space. Non-gaussian distributions required different fitting algorithms to parameterize the beam shape. Since a single measurement required many beam pulses, it was essential to filter out errant data. Beam position monitors near the scanners were used to fit the trajectory on each pulse and correct the expected position of the beam with respect to the wire. Automated procedures require robust fitting algorithms with careful error analysis. The accumulated SLC experience underscores several essential requirements for linear collider diagnostics. In addition to providing sufficient precision, the scans must be non-invasive to allow frequent measurements during normal operation. Automated procedures are needed so the scans can be regularly scheduled to provide long term history and allow correlation with other events. Future collider designs have incorporated these lessons and included precision diagnostics and correction elements.

## 5 FEEDBACK

Another lesson from the SLC experience is the crucial importance of feedback to combat the inherent instabilities of a linear collider. Feedback controlled the beam energy and trajectory, stabilized the polarized source, and maintained and optimized collisions. Several generations of development were required to produce the flexible feedback systems used throughout the SLC. The first 'slow' SLC energy and trajectory feedback was implemented in 1985. This was followed by prototype pulse-to-pulse feedback using dedicated hardware. Energy and trajectory feedback at the end of the linac was developed in 1987 and collision feedback in 1989. A generalized database-driven system [11] was implemented starting in 1991. This feedback used existing hardware, making it relatively easy to add a new system anywhere needed. In order to avoid overcorrection, the sequence of trajectory feedbacks along the main linac were connected by a 'cascade' system which allowed each feedback to communicate with its next downstream neighbor. Transfer matrices between the feedbacks were adaptively calculated. Online diagnostics of the feedback performance were expanded over several years to provide better monitoring and histories. A luminosity optimization feedback was developed in 1997 to improve the resolution of the final optical tuning at the IP. By 1998, the SLC had more than 50 feedback systems controlling over 250 beam parameters.

The optimization feedback is an interesting example of a system which may have wide applicability for future machines. To achieve and maintain the minimum beam size at the SLC IP, five final corrections were routinely optimized for each beam. These included centering of the $x$ and $y$ beam waist positions, zeroing of the dispersion $\eta_x$ and $\eta_y$, and minimization of an $x$–$y$ coupling term. Since the first SLC collisions, an automated procedure was used to scan the beam size as a function of each parameter and set the optimum value. The beam size was measured with a beam-beam deflection scan but this technique lacked the resolution required to accurately measure micron-size, disrupted beams. It was estimated that poor optimization caused a 20-30% reduction in luminosity during the 1996 run [12]. For 1997, a novel 'dithering' feedback was implemented which optimized a direct measure of the luminosity (i.e. the beamstrahlung signal) as a function of small changes in each parameter [13]. By averaging over 1000s of beam pulses, it was possible to improve the resolution by a factor of 10. A similar 'dithering' feedback was developed to minimize emittance at the end of the linac but never fully commissioned. Optimization feedback modeled on this system has been incorporated in the designs of future colliders.

The SLC feedback systems were essential for reliable operation of the accelerator and provided several less obvious benefits. Feedback compensated for slow environmental changes such as diurnal temperature drifts or decreasing laser intensity and provided a fast response to changes such as klystrons cycling. It facilitated a smooth recovery from any interruption to operation. Feedback improved operating efficiency by providing uniform performance independent of the attention or proficiency of a particular operations crew. An important benefit was that the feedback decoupled different systems so that tuning could proceed non-invasively in different parts of the machine while delivering luminosity. Feedback also provided a very powerful monitor of many aspects of the machine performance. Much of the SLC progress came from using feedback to automate as many routine tuning operations as possible.

In spite of the critical importance of feedback for SLC operation, there were several areas in which the feedback performance was less than optimal. Many of the problems occurred in the sequence of trajectory feedbacks along the main linac. To avoid having multiple feedbacks respond to an incoming disturbance, a simple one-to-one system was used where each loop communicated with the next downstream feedback. The topology of this 'cascade' system was limited by bandwidth and connectivity constraints. However, in the presence of strong wakefields, the beam transport depends on the origin of the perturbation and a more complex interconnection is required. Each feedback must have information from all upstream systems to determine the ideal orbit correction. Simulations indicate that a feedback system with a many-to-one cascade can avoid overcorrection problems [14]. Another problem which has been studied and understood concerns the configuration of monitors and correctors in each system. Traditional SLC feedbacks were distributed over at most two sectors (200 m), constraining the orbit locally but allowing oscillations to grow elsewhere. Simulations have shown that a better trajectory can be achieved if the devices are distributed over a longer region. The success of the SLC feedback systems coupled with new understanding of their limitations has produced a robust design for the feedback required for the NLC.

## 6 TUNING ALGORITHMS

A variety of innovative optical tuning techniques were developed for the SLC. These included precision beam-based alignment using ballistic data and other methods, as well as betatron and dispersion matching. The luminosity optimization feedback provided the resolution required to align the final focus sextupoles and octupoles. In the non-planar SLC arcs, a 4-D transfer matrix reconstruction technique with careful error analysis allowed minimization of coupling terms and synchrotron radiation emittance growth. An extension of this method allowed an adjustment of the effective spin tune of the arc to preserve maximum polarization. A two-beam dispersion free steering algorithm developed for the SLC linac was later applied successfully at LEP.

An important technique used for emittance control in the SLC linac was to introduce a deliberate betatron oscillation to generate wakefield tails which compensated for those due to alignment errors [15]. Wire scanner measurements of the beam profile were used to characterize the wakefield tails, and then an oscillation was created by one of the linac trajectory feedbacks which was closed by the next feedback. Since 1991, this method was applied with reasonable success using wires in the middle and near the end of the linac. One problem was that careful tuning was required to find the optimal phase and amplitude for the oscillation. The cancellation is also very sensitive to the phase advance between the source of the wakefield and the compensating oscillation so any change in the optics required retuning. Simulations also showed that significant emittance growth could occur in the 200 m of linac downstream of the last wires [16]. For the 1997 run, a different strategy was adopted. Wire scanners at the entrance to the final focus were used for tuning out wakefield tails to ensure that the entire linac was compensated. In addition, the induced oscillations were made nearer the end of the linac where the higher energy beam was less sensitive to optics changes, making the tuning more stable. This technique was successful in reducing the emittance growth in the SLC linac by more than an order of magnitude. Figure 6 shows the evolution of beam size over time.

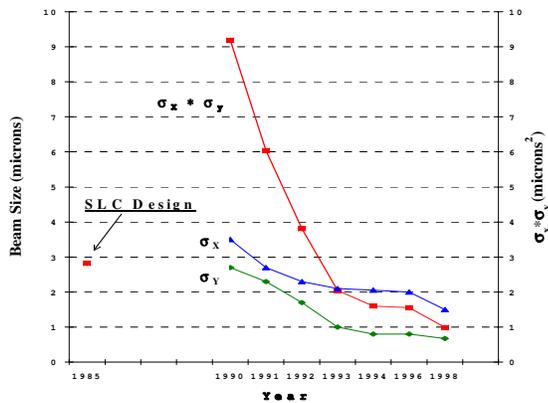

Figure 5: Evolution of beam size at the SLC IP over time from 1991 to 1998 showing $\sigma_x$ (blue-middle), $\sigma_y$ (green-lower) and the product $\sigma_x\sigma_y$ (red-upper). The final beam area was 1/3 of the original design value.

## 7 CONCLUSIONS

More than ten years of SLC operation has produced much valuable experience for future linear colliders. Because a linear collider lacks the inherent stability of a storage ring, it is a much more difficult machine. Significant progress was made on precision diagnostics for beam characterization and on flexible, intelligent feedback systems. New techniques for optical matching, beam-based alignment, and wakefield control were developed and refined. The beam-beam deflection was shown to be a powerful tool for stabilizing and optimizing collisions. Important lessons were also learned on collimation and background control and on many other issues not discussed here. Both the SLC and future colliders benefited from an intense exchange of ideas and experiments. It is the experience and knowledge gained with the SLC that gives confidence that the NLC design contains the tools required to commission and operate a linear collider.

The most enduring lesson from the SLC is undoubtedly that any new accelerator technology will present unanticipated challenges and require considerable hard work to master. Once a technology becomes routine, it is easy to forget the initial effort that was needed. At the SLC as elsewhere, the most difficult problems were almost always those which were not expected. It is also clear that the experience gained on an operating accelerator is complementary to that from demonstration projects. The discipline of trying to produce physics forces one to confront and solve problems which are not relevant otherwise. The SLC finally reached near design luminosity due to the creativity and dedication of a large number of people over many years.